\documentclass[12pt]{article}

\usepackage{epsf}
\usepackage{amsmath,amssymb}
\usepackage{latexsym}
\usepackage{wrapfig}
\usepackage{latexsym}
\usepackage[dvipdfm]{graphicx}
\usepackage{color}
\usepackage{delarray,color}
\usepackage{fancybox}  

\newcommand {\beq}{\begin{eqnarray}}
\newcommand {\eeq}{\end{eqnarray}}

\newcommand{\be}{\begin{equation}}
\newcommand{\ba}{\begin{eqnarray}}
\newcommand{\ea}{\end{eqnarray}}
\newcommand{\ee}{\end{equation}}

\newcommand{\z}{{\mathbf Z }}

\newcommand{\beqa}{\begin{eqnarray}}
\newcommand{\eeqa}{\end{eqnarray}}

\newcommand{\unit}{\hbox to 3.8pt{\hskip1.3pt \vrule height 7.4pt
    width .4pt \hskip.7pt \vrule height 7.85pt width .4pt \kern-2.4pt
    \hrulefill \kern-3pt \raise 3.7pt\hbox{\char'40}}}

\def\matt[#1,#2,#3,#4]{\left(%
\begin{array}{cc} #1 & #2 \\ #3 & #4 \end{array} \right)}

\def\d{{\rm d}}

\def\dg{\dagger}
\def\Im{{\rm Im}}
\def\max{{\rm max}}
\def\min{{\rm min}}
\def\Re{{\rm Re}}

\def\til{\widetilde}
\usepackage{tabularx}

\catcode`\@=11
\@addtoreset{equation}{section}

\catcode`@=12
\relax

\newcommand{\tr}{\mathrm{Tr}}

\begin{document}

\begin{titlepage}

\setcounter{page}{0}

\renewcommand{\thefootnote}{\fnsymbol{footnote}}

\begin{flushright}
YITP-12-64\\
\end{flushright}

\vskip 1.35cm

\begin{center}
{\Large \bf 
M5-branes in ABJM theory and Nahm equation
}

\vskip 1.2cm 

{\normalsize
Tomoki Nosaka\footnote{nosaka(at)yukawa.kyoto-u.ac.jp} and Seiji Terashima\footnote{terasima(at)yukawa.kyoto-u.ac.jp}
}

\vskip 0.8cm

{ \it
Yukawa Institute for Theoretical Physics, Kyoto University, Kyoto 606-8502, Japan
}

\end{center}

\vspace{12mm}

\centerline{{\bf Abstract}}

\vskip 0.5cm 

We explicitly construct two classes of the BPS solutions in the ABJM action: the funnel type solutions and the 't Hooft-Polyakov type solutions, and 
study their physical properties as the M2-M5 bound state.
Furthermore, we give a one to one correspondence between the solutions of the BPS equation and the ones of an extended Nahm equation which 
includes the Nahm equation.
This enables us to construct infinitely many conserved quantities from the Lax form of the Nahm equation.

\end{titlepage}
\newpage

\tableofcontents
\vskip 1.2cm 

\section{Introduction and Summary}

After the ground-breaking work by Bagger and Lambert \cite{BLG1} and Gustavsson \cite{BLG2}, the 
multiple M2-branes have been studied intensively 
and a three dimensional ${\cal N}=6$ supersymmetric Chern-Simons-matter conformal field theory with gauge group $U(N) \times U(N)$ was 
proposed as an action of the low energy limit of $N$ M2-branes on ${\mathbf C}^4 / \z_k$ by 
Aharony-Bergman-Jafferis-Maldacena (ABJM) \cite{ABJM}.
There have been significant progress in understanding the M2-branes by the ABJM action. (See \cite{review} for the recent review of this subject.)

On the other hand, the M5-branes have been still poorly understood.
The theory on the multiple M5-branes are highly mysterious.
For instance, the gravity dual analysis implies that there should be ${\cal O}(N^3)$ degrees of freedom at large $N$, which can not be understood 
from gauge theory.\footnote{
Recently, it was claimed \cite{Minahan, SKim} that the ${\cal O}(N^3)$ behavior is reproduced from the localization computation of the 5d SUSY 
gauge theory on $S^5$ \cite{HST} \cite{Kallen1, Kallen2}.
}
In order to study the M5-branes, the ABJM action will be useful because the bound state of the M5-branes and the M2-branes can be described by 
the M2-brane action, where the M5-branes will be the "solitons" of the action.
Indeed, the BPS solution corresponding to the funnel type bound state of these was found in \cite{Te, GRVV}, which can be regarded as a variant of 
the famous solution in the BLG action by Basu and Havey \cite{BaHa}, and have been studied further \cite{Hanaki,Nastase:2009ny, Gus}.\footnote{
Moreover, an M-theory lift of the D4-branes with a constant magnetic field in type IIA string theory should be an M2-M5 bound state and it was also 
constructed in the ABJM action \cite{TY1,TY2}.
This system is also useful for understanding the M5-branes.
}
This is the M-theory lift of the bound state of the D2-branes and the D4-branes which are described as the solution of the Nahm equation from the 
D2-branes or the monopole from the D4-branes.
The shape of the solution should be a fuzzy $S^3/\z_k$ at a point in the world volume of the M2-branes.

For the D2-D4 bound state (which is essentially same as the D1-D3 bound state \cite{Diaco}), we can use the Nahm construction 
\cite{Corrigan-Goddard} to construct the monopole solution from the Nahm data.
For the M2-M5 bound state, we expect that there will be such correspondence between the BPS solution in the multiple M5-branes and the ones in 
the ABJM action.\footnote{
For the BLG action, Gustavsson studied this \cite{Gustavsson}.
In \cite{Sae1, Sae2}, the Nahm construction for the BLG and the ABJM actions were considered, but they assume rotational invariance of the 
solutions.
}
This may give us some important clues for understanding the M5-branes.\footnote{
We cannot use the BLG theory instead of the ABJM theory.
Since scalar fields in the BLG theory live in ${\cal A}_4$, there 
are no natural way to get from the BPS solution the information of the positions of the M2-branes which is necessary to discuss how that bound 
state should be expressed on the M5-branes.
}
For this project, we obviously need the details of the solutions of the BPS equation of the ABJM action.
However, the solutions of the BPS equations \cite{Te} are less known and the properties of the solutions, for example what is the moduli space, 
have not been studied.
Even the positions of the M2-branes far from the M5-branes are unclear and ambiguous, as we see in section 2.

In this paper, we construct two classes of the BPS solutions explicitly, and study their physical properties.\footnote{
One can obtain further BPS solutions by taking the direct sum of these BPS solutions.
One can even construct the bound state of the M5-branes, each of which extends in different directions in space-time, and the M2-branes 
\cite{Krishnan}.
Though we do not consider in this paper, this direction would also be interesting as a future work.
}
Solutions in one class include the one found in \cite{Te, GRVV}, but the positions of the M2-branes are more general.
Solutions in the other class behave like the Nahm data of the 't Hooft-Polyakov monopole and represent the bound state of two M5-branes.
Furthermore, we give a one to one correspondence between the solutions of the BPS equation and the ones of an extended Nahm equation which 
includes the Nahm equation.
This enables us to construct infinitely many conserved quantities from the Lax form of the Nahm equation.
We also investigate the space-time profiles of the solutions using the correspondence.

The organization of this paper is as follows.
In section two we construct the BPS solutions representing the M2-branes ending on the M5-branes and investigate their profiles in space-time.
In section three we show the one to one correspondence between the BPS solutions and the extended Nahm data, and using this, construct 
conserved quantities of the BPS solution.
We also comment on the relation to the reduction from the M2-branes to the D2-branes discussed by Mukhi and Papageorgakis \cite{Mukhi}.

\section{The BPS solutions representing the M2-branes ending on the M5-branes}

In this paper, we study the half BPS solutions of the ABJM action which represent the M2-branes ending on the M5-branes.
We assume that the M2-branes extend in $(x^0,x^5,x^6)$ directions and that the fields on the M2-branes depend on $x^6$, which we will denote 
$s$, only.
The bosonic fields of the ABJM action are the gauge fields and the $N \times N$ matrix valued complex scalar fields $Y^1,Y^2,Y^3,Y^4$ 
representing the positions in the transverse directions of M2-branes.
Here $N$ is the number of the M2-branes.
We also assume that $Y^3, Y^4$ and the gauge fields vanish in the BPS solution.
This implies that the M5-branes are extending along $(x^0,x^1,x^2,x^3,x^4,x^5)$ where we identify the directions of $(Y^1,Y^2)$ with 
$(x^1,x^2,x^3,x^4)$.

Thus the BPS equation \cite{Te} \cite{GRVV} is 
\begin{align}	
\dot{Y}^a={2\pi\over k}(Y^aY^{b\dg}Y^b-Y^bY^{b\dg}Y^a),
\label{theBPS}
\end{align}
where $k$ is the level of the Chern-Simons action, $Y^a$ $(a=1,2)$ is the $N \times N$ matrix valued scalar field representing the positions in the 
transverse directions of M2-branes and $\dot{Y}^a= \frac{d Y^a}{d s}$.
In the type IIA limit, the M2-M5 bound state reduces to the D2-D4 bound states, which is described by the Nahm equation from the D2-brane 
viewpoint and by the monopole equation from the D4-brane viewpoint.
Therefore, the BPS equation (\ref{theBPS}) is an analogue of the Nahm equation in the M-theory.
Note that the r.h.s. of (\ref{theBPS}) can be written by the Lie 3-algebra \cite{BL} and then the equation (\ref{theBPS}) can be regarded as a 
generalization of the Basu-Harvey equation \cite{BaHa} to the ABJM action.

Note that (\ref{theBPS}) is covariant under the $U(N)\times U(N)$ gauge transformation, $Y^a\rightarrow UY^aV^\dg$, and the $SU(2)$ global 
transformation, $Y^a\rightarrow\Lambda^{ab}Y^b$, of the ABJM action, where $U, V$ should be constant because of our assumption 
$A_\mu=\tilde{A}_\mu=0$.

\subsection{The funnel type solutions}\label{sfs}

Here we will explicitly construct the solutions of the BPS equation (\ref{theBPS}) which represent $N$ M2-branes ending on a M5-brane at $s=s_0$ 
and extending to $s=\infty$.
We take the following ansatz for the solutions:
\begin{align}
Y^a(s)=\sqrt{k\over 4\pi}f_a (s) G^a,
\end{align}
where $f_a(s)$ is a function of $s$ and $G^a$ is the constant $N \times N$ matrix defined by \cite{Te} \cite{GRVV}
\begin{align}
-G^a=G^a{G^b}^\dg G^b-G^b{G^b}^\dg G^a.
\end{align}
Using the $U(N) \times U(N)$ gauge symmetry of the ABJM action, we can express them as
\begin{align}
G^1_{m n}&=\delta_{m,n-1}\sqrt{m},\nonumber\\
G^2_{m n}&=\delta_{m,n}\sqrt{N-n}\label{Ga},
\end{align}
where $m,n =1, \ldots,N$.
Then the BPS equation reduces to\footnote{
This solution was considered in \cite{Te} and also in a recent 
work \cite{Mohammed:2012rd} independently.
}
\begin{align}
\dot{f_a}=-{1\over 2}f_a|f_b|^2(b\neq a).
\end{align}
By the symmetry of the BPS equation, we can replace
\begin{align}
Y^a\rightarrow\Lambda^{ab}UY^bV^\dg.\label{symmetrytrsf}
\end{align}
This can make $f_a$ satisfy $f_2\ge f_1\ge 0$.
Then, we can write down the solution explicitly:
\begin{align}
Y^1(s)&=\sqrt{k\over 4\pi}G^1\cdot{C\exp[-C^2(s-s_0)/2]\over\sqrt{1-\exp [-C^2(s-s_0)]}},\nonumber\\
Y^2(s)&=\sqrt{k\over 4\pi}G^2\cdot{C\over\sqrt{1-\exp[-C^2(s-s_0)]}},\label{shifted}
\end{align}
where
\begin{align}
C^2=(f_2)^2-(f_1)^2
\end{align}
is a constant.

\begin{figure}[htbp]
 \begin{center}
  \includegraphics[width=100mm]{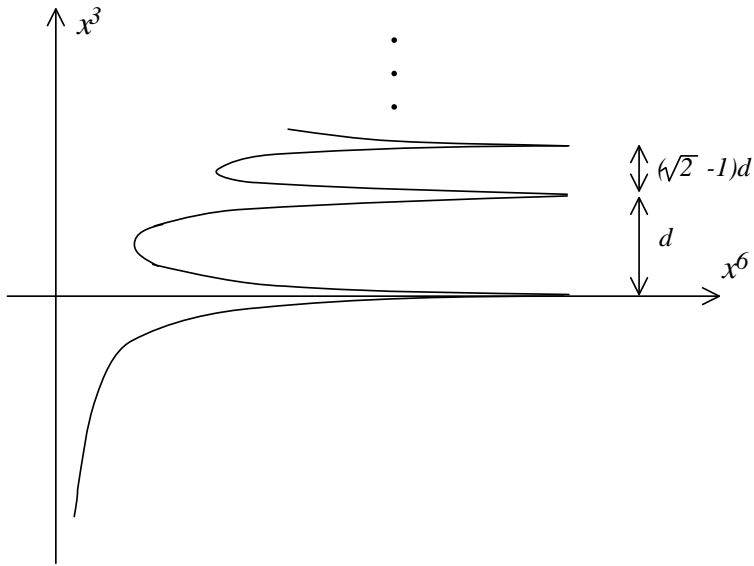}
\end{center}
\caption{The expected profile of the solution, where $d=C\sqrt{k\over 4\pi}$. 
\label{fig1}}
\end{figure}

Now we will study the physical interpretation of this solution.
First, $s_0$ is the position of the M5-brane because $Y^a$ diverges at $s=s_0$ as in the solution obtained in \cite{Te,GRVV}.
We will shift the coordinate $s$ such that $s_0=0$.
If $Y^{1,2}(s)$ are equivalent to diagonal matrices by $U(N)\times U(N)$, we expect that the $i$-th eigenvalues of $Y^a$ 
represent the position of the $i$-th M2-brane.
Here defining $z^1\equiv x^1+ix^2,z^2\equiv x^3+ix^4$, we identify the eigenvalues of $Y^a$ with $z^a$.\footnote{
This diagonalization has $U(1)^N$ ambiguity, i.e. 
$z^a\rightarrow{\rm e}^{i\theta}z^a$ for each diagonal component.
However, we expect them physically inequivalent due to Chern-Simons 
term, in the same way to the analysis of vacuum moduli space in \cite{ABJM}
}
Then, at $s=\infty$, the position of the $i$-th M2-brane ($i=1,\ldots,N$) is
\begin{align}
\begin{bmatrix}x^1\\
x^2\\
x^3\\
x^4
\end{bmatrix}
=
\begin{bmatrix}0\\
0\\
C \sqrt{k(N-i)\over 4\pi}\\
0
\end{bmatrix},
\end{align}
which is shown in Fig.~\ref{fig1}.
This clearly shows that the solution is not spherically symmetric.
Thus, the symmetry rotations gives different solutions.
For example, using the global $SU(2)$ and a choice of $C$, we can set an M2-brane at an arbitrary point in $(z^1,z^2)$.
Furthermore, because the scalars are diagonalized at $s=\infty$, we can make an $U(1)^N$ transformation, which is the 
diagonal part of $U(N)$, to move each of the M2-branes at $z_i^a$ to ${\mathrm e}^{i\phi_i}z_i^a$.

Note that for the funnel type solution obtained in \cite{Te,GRVV},
\begin{align}
Y^a(s)&=\sqrt{k\over 4\pi} G^a \frac{1}{\sqrt{s-s_0}},
\label{t1}
\end{align}
all the M2-branes are at $z^1=z^2=0$ at $s=\infty$.
Indeed, our solution contains this as the $C\rightarrow 0$ limit.
Moreover, we can easily see that if $Y^a$ diverges at a point $s=s_0$, the solution should be approximated by a diagonal 
sum of the solutions (\ref{t1}) with a symmetry transformation (\ref{symmetrytrsf}).
This means there is an M5-brane at $s=s_0$ because of the interpretation of the solution (\ref{t1}).
We can check our solution behave like this near $s=s_0$.

\subsection{'t Hooft Polyakov type solutions}\label{tH}

In this subsection, we will consider $N=2$ case only and construct a solution of the BPS equation corresponding to the 
Nahm data of the 't Hooft Polyakov monopole.
We take the following ansatz:
\begin{align}	
Y^1(s)&=\sqrt{k\over 4\pi}(f_1(s) \sigma^1+if_2(s) \sigma^2),\nonumber\\
Y^2(s)&=\sqrt{k\over 4\pi}(f_3(s) \sigma^3+f_4(s)\sigma^4),
\label{tH1}
\end{align}
where $f_i(s)$ is real and $\sigma^1,\sigma^2,\sigma^3$ are Pauli matrices and $\sigma^4$ is the unit matrix.\footnote{
A solution of the similar form for the Basu-Harvey equation was found 
in \cite{Nogradi:2005yk}.
The BLG action for the $A_4$ algebra is equivalent to the $SU(2) \times SU(2)$ 
ABJM action which is equivalent to the $N=2$ ABJM action if we forget the gauge fields.
Thus our solution and the one in \cite{Nogradi:2005yk} are essentially 
same by an appropriate map. 
In BLG theory, however, we did not know the relation between the 
parameters and the positions of the M2-branes. 
}

\begin{figure}[htbp]
 \begin{center}
  \includegraphics[width=100mm]{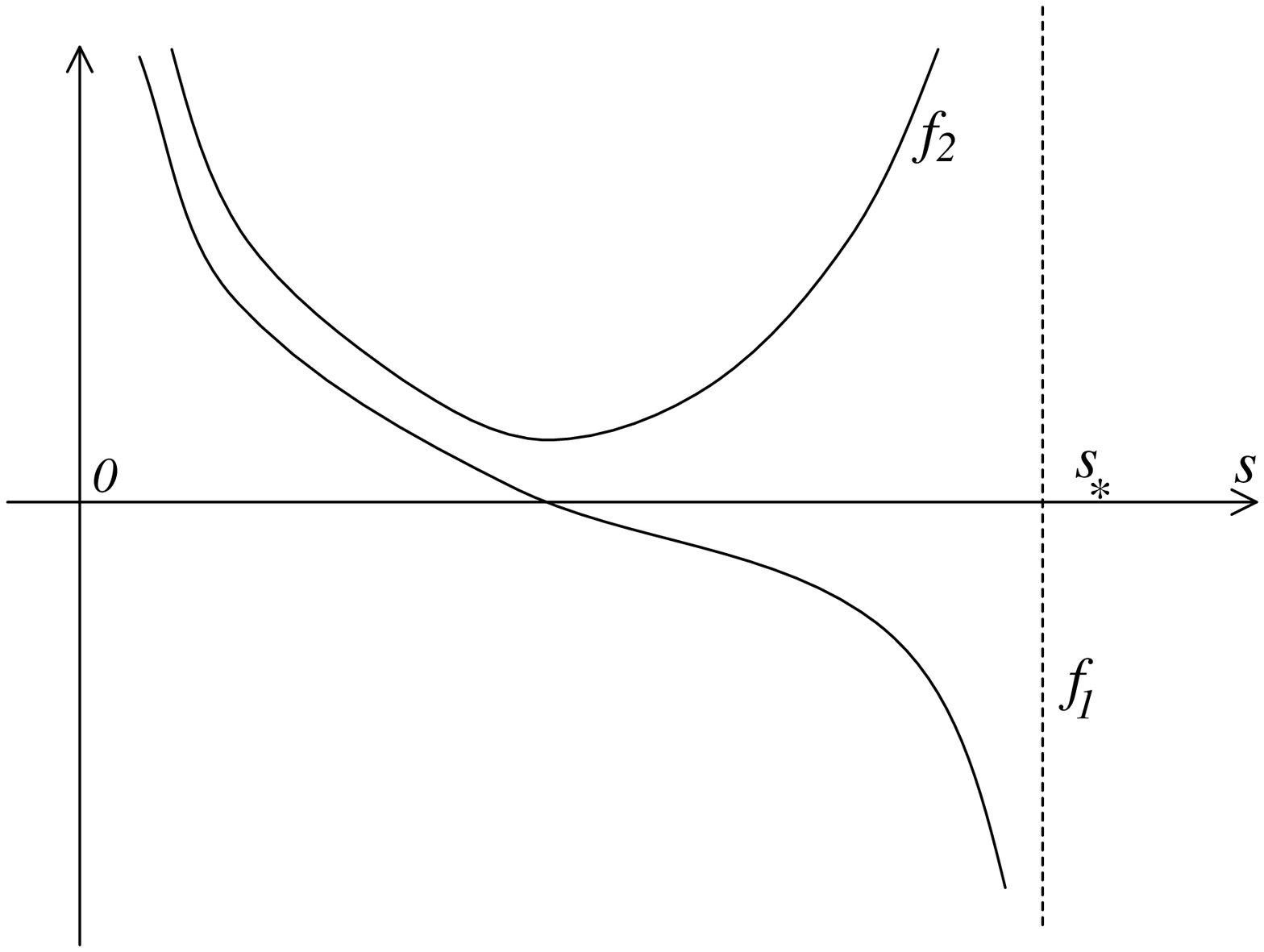}
 \end{center}
\caption{Profiles of $f_1$ and $f_2$
\label{fig2}}
\end{figure}

By the symmetry transformation (\ref{symmetrytrsf}),
we can make $|f_1|\le f_{2,3,4}$ at a given point in $s$, say $s=s_0$.
Then, the BPS equation becomes
\begin{eqnarray}
\dot{f_i}=-2 f_j f_k f_l,
\end{eqnarray}
where $\epsilon_{ijkl} \neq 0$.
This implies there are following independent conserved quantities,
\begin{align}
\alpha^2&\equiv f_2^2-f_1^2,\nonumber\\
\beta^2&\equiv f_3^2-f_1^2,\nonumber\\
\gamma^2&\equiv f_4^2-f_1^2\label{alphabetagamma}.
\end{align}
The remaining equation is 
\begin{align}
\dot{f_1}&=-2 \sqrt{(f_1^2+\alpha^2)(f_1^2+\beta^2)(f_1^2+\gamma^2)},
\end{align}
which is solved as
\begin{align}
2 s&=\int^\infty_{f_1(s)}{d f\over\sqrt{(f^2+\alpha^2)
(f^2+\beta^2)(f^2+\gamma^2)}},\label{theintegrationoff1}\\
&f_2(s)=\sqrt{\alpha^2+f_1^2},\nonumber\\
&f_3(s)=\sqrt{\beta^2+f_1^2},\nonumber\\
&f_4(s)=\sqrt{\gamma^2+f_1^2},\label{tHooftsolution}
\end{align}
where we have chosen the integration constant such that 
$f_1(s)=\infty$ at $s=0$.
We can see that $f_1$ also diverges at $s=s_*$, where 
\begin{align}
s_\ast={1\over 2}\int^\infty_{-\infty}{d f\over\sqrt{(f^2+\alpha^2)(f^2+\beta^2)(f^2+\gamma^2)}}.\label{s*int}
\end{align}
Thus, this solution represents the two M2-branes stretching between the two M5 branes at $s=0$ and $s=s_\ast$.
Using the first kind of the elliptic integral
\begin{align}
{\cal F}(\phi,k)\equiv\int^\phi_0{d \theta\over\sqrt{1-k^2\sin^2\theta}}\label{elliptic},
\end{align}
the integration (\ref{theintegrationoff1}) is written as
\begin{align}
\left|{s_\ast\over 2}-s\right| =& 
{1\over
 2\sqrt{\alpha_2^2(\alpha_1^2-\alpha_3^2)}}{\cal
 F}\left(\arcsin\sqrt{{\alpha_1^2-\alpha_3^2\over\alpha_1^2}{f_1^2\over
 f_1^2+\alpha_3^2}},\sqrt{\alpha_1^2(\alpha_2^2-\alpha_3^2)\over\alpha_2^2(\alpha_1^2-\alpha_3^2)}\right) \nonumber\\
=& 
{1\over
 2\sqrt{\alpha_2^2(\alpha_1^2-\alpha_3^2)}}
\mathrm{sn}^{-1}
\left(\sqrt{{\alpha_1^2-\alpha_3^2\over\alpha_1^2}{f_1^2\over
 f_1^2+\alpha_3^2}},\sqrt{\alpha_1^2(\alpha_2^2-\alpha_3^2)\over\alpha_2^2(\alpha_1^2-\alpha_3^2)}\right),
\label{f1}
\end{align}
where
\begin{align}
s_\ast &={1\over\sqrt{\alpha_2^2(\alpha_1^2-\alpha_3^2)}}{\cal F}\left(\arcsin\sqrt{\alpha_1^2-\alpha_3^2\over\alpha_1^2},\sqrt{\alpha_1^2(\alpha_2^2-\alpha_3^2)\over\alpha_2^2(\alpha_1^2-\alpha_3^2)}\right)\label{s*ell}
\end{align}
and 
$(\alpha_1,\alpha_2,\alpha_3)$ is $(\alpha,\beta,\gamma)$ or another permutation such that $\alpha_1^2 \geq \alpha_2^2 \geq \alpha_3^2$.
The schematic profiles of $f_1$ and $f_2$ are shown in Fig.~\ref{fig2}.

For $\alpha=0$, we find 
\begin{align}
s_\ast=\int^\infty_0{\d f\over
f\sqrt{(f^2+\beta^2)(f^2+\gamma^2)}}=\infty,
\end{align}
which means there is only one M5-brane and the solution is in funnel shape.
In this case, because $f_1(\infty)=f_2(\infty)=0$, $f_3(\infty)=\beta$ and $f_4(\infty)=\gamma$, the positions of the two M2-branes are 
\begin{align}
\begin{bmatrix}
x^1\\
x^2\\
x^3\\
x^4
\end{bmatrix}=
\begin{bmatrix}
0\\
0\\
\sqrt{k\over 4\pi}(\beta+\gamma)\\
0
\end{bmatrix}, \,\,
\begin{bmatrix}
x^1\\
x^2\\
x^3\\
x^4
\end{bmatrix}=
\begin{bmatrix}
0\\
0\\
\sqrt{k\over 4\pi}(-\beta+\gamma)\\
0
\end{bmatrix}.
\end{align}
By the choice of $\beta$, $\gamma$ and the symmetry transformation (\ref{symmetrytrsf}) with 
\begin{align}
&\beta=\sqrt{\pi\over k}(R_1+R_2), \,\,\,
\gamma=\sqrt{\pi\over k}(R_1-R_2),\\
&\Lambda=
\begin{pmatrix}
\sin\theta_1{\rm e}^{i(\phi_1-\psi_1)} &\cos\theta_1\\
-\cos\theta_1 &\sin\theta_1{\rm e}^{i(\psi_1-\phi_1)}
\end{pmatrix}, \,\,\,
U=
\begin{pmatrix}
{\rm e}^{i\phi_1} &0\\
0 &{\rm e}^{i\phi_2}
\end{pmatrix}, \,\,\,
V=1,
\end{align}
we can have the solution representing the two M2-branes at
\begin{align}
\begin{bmatrix}
x^1\\
x^2\\
x^3\\
x^4
\end{bmatrix}=
\begin{bmatrix}
R_1\cos\theta_1\cos\phi_1\\
R_1\cos\theta_1\sin\phi_1\\
R_1\sin\theta_1\cos\psi_1\\
R_1\sin\theta_1\sin\psi_1
\end{bmatrix}, \,\,\,\,
\begin{bmatrix}
x^1\\
x^2\\
x^3\\
x^4
\end{bmatrix}=
\begin{bmatrix}
R_2\cos\theta_1\cos\phi_2\\
R_2\cos\theta_1\sin\phi_2\\
R_2\sin\theta_1\cos(\psi_1+\phi_2-\phi_1)\\
R_2\sin\theta_1\sin(\psi_1+\phi_2-\phi_1)
\end{bmatrix}.
\end{align}
If we further put $\beta=\gamma$, the solution reduces to the funnel type solution (\ref{shifted}) obtained in the previous subsection for $N=2$. 

Finally, we will consider the limit of the reduction from the M2-branes to the D2-branes discussed by Mukhi and Papageorgakis for the solution (\ref{tH1}).
We take the parameters as
\begin{align}
\alpha={\alpha_0\over \sqrt{4\pi k}},\,\,\,
\beta={\beta_0\over\sqrt{4\pi k}},\,\,\,
\gamma=\gamma_0\sqrt{k\over 4\pi},\label{Mukhiabg}
\end{align}
and take the $k\rightarrow\infty$ limit with 
$\alpha_0$, $\beta_0$ and $\gamma_0$ fixed.
We assume $\beta_0>\alpha_0$ for simplicity.
In this case, by (\ref{s*ell}) (here $(\alpha_1,\alpha_2,\alpha_3)=(\gamma,\beta,\alpha)$), 
\begin{align}
s_*&={4\pi\over\sqrt{\beta_0^2 (\gamma_0^2-{\alpha_0^2\over k^2})}}{\cal F}\left(\arcsin\sqrt{1-{1\over k^2}\cdot {\alpha_0^2 \over \gamma_0^2}},\sqrt{\beta_0^2-\alpha_0^2\over\beta_0^2(1-{1\over k^2}\cdot {\alpha_0^2 \over \gamma_0^2})}\right)\nonumber\\
&\hspace{-2mm}\underset{k\rightarrow\infty}{\rightarrow}
{4\pi\over \beta_0 \gamma_0}{\cal F}\left(\arcsin(1),\sqrt{1-{\alpha_0^2\over\beta_0^2}}\right)
\end{align}
is finite.
Also, around $s=s_*/2$, since $f_1\approx 0,f_2\approx \alpha,f_3\approx\beta$ and $f_4\approx\gamma$, 
\begin{align}
Y^1&=\mathcal{O}(1),\\
Y^2&={k\over 4\pi}\gamma_0+\mathcal{O}(1).
\end{align}
The region around $s_*/2$ where these are true is of finite range.
Indeed, the integration (\ref{s*int}) is dominated by the contribution from $f=\mathcal{O}(1/\sqrt{k})$.
Therefore, by (\ref{theintegrationoff1}), 
if $|s-s_*|=\mathcal{O}(1)$, then $f_1(s)=\mathcal{O}(1/\sqrt{k})$.
With $\gamma_0=4\pi v/k$, this behaviour of $Y^a$ is just as assumed in the reduction from the M2-branes to the D2-branes 
(\ref{MukhiT}) 
discussed in section 3.

\section{Nahm data and ABJM}

In this section, we will show that any solution of the BPS equation (\ref{theBPS}) of the ABJM action gives two 
sets of Nahm data of $N$ monopoles which satisfy the Nahm equation.
This correspondence allow us to construct the conserved quantities.

In the ABJM action, there is the $U(N) \times U(N)$ gauge symmetry.
By taking the product of $Y^a$ and ${Y^a}^\dg$ we have a adjoint representation of a $U(N)$ gauge symmetry which 
is a singlet under the other $U(N)$.
We will define
\begin{align}
T^M(s)=&{2\pi\over k}\sigma^M_{ab}Y^b{Y^a}^\dg\label{TMs}
\end{align}
and
\begin{align}
\til{T}^M(s)=&{2\pi\over k}\bar{\sigma}^M_{ab}{Y^a}^\dg Y^b,\label{tTMs}
\end{align}
where
\begin{align}
\sigma^M=(\sigma^I,1),\,\,\,\bar{\sigma}^M=(\sigma^I,-1),
\end{align}
$M=1and \ldots,4$, $I=1,2,3$.
Under the $U(N)\times U(N)$ transformation ($Y^a\rightarrow UY^aV^\dg$), they transform as
\begin{align}
T^M\rightarrow UT^MU^\dg,\,\,\,\til{T}^M\rightarrow V\til{T}^MV^\dg.
\end{align}
Then, from the BPS equation of the ABJM theory (\ref{theBPS}) we can show that $T^M$ and $\til{T}^M$ satisfy the 
following differential equations: 
\begin{align}	
\dot{T^I}&=i\epsilon_{IJK}T^JT^K,\label{eq:TM1}\\
\dot{T^4}&=(T^1)^2+(T^2)^2+(T^3)^2-(T^4)^2,\label{eq:TM2}
\end{align}
and
\begin{align}	
\dot{\til{T}^I}&=i\epsilon_{IJK}\til{T}^J\til{T}^K,\label{eq:TM3}\\
\dot{\til{T}^4}&=(\til{T}^1)^2+(\til{T}^2)^2+(\til{T}^3)^2-(\til{T}^4)^2.\label{eq:TM4}
\end{align}
Therefore, $T^I$ ($\til{T}^I$) is the solution of the Nahm equation (\ref{eq:TM1}) ((\ref{eq:TM3})) which 
is called Nahm data with appropriate boundary conditions.\footnote{
This can be considered as a generalization of the result for the 
BLG action in \cite{Gustavsson}.
A similar phenomena for the monopole in the ABJM action was 
observed in \cite{Hosomichi}.
}
Here, we have four matrices which are the Nahm data $T^I$ ($\til{T}^I$) and an additional one 
$T^4$ ($\til{T}^4$).
We will call them extended Nahm data and the differential equations (\ref{eq:TM1}), (\ref{eq:TM2})
((\ref{eq:TM3}), (\ref{eq:TM4})) the extended Nahm equation.

Now we will show the one to one correspondence 
\begin{align}
\{Y^a(s)\}/V\longleftrightarrow\{T^M(s)|T(s_0)=AA^\dg\text{ for some }A\in Mat_{2N\times N}(\mathbf{C})\},\label{onetoone} 
\end{align}
where we defined
\begin{align}
T\equiv\sigma^M\otimes T^M=
\begin{bmatrix}
T^4+T^3 &T^1-iT^2\\
T^1+iT^2 &T^4-T^3
\end{bmatrix}.
\end{align}
$s_0$ is a fixed constant and we consider the BPS solutions and the extended Nahm data which are finite 
in the neighbourhood of $s_0$.

First, the quotient in the left hand side is necessary since $(Y^1,Y^2)$ and $(Y^1V,Y^2V)$ give the same $T^M$ by (\ref{TMs}).
This $V$ must be a constant matrix since a local transformation is forbidden by the gauge fixing in (\ref{theBPS}).
Second, since the extended Nahm equation is the set of the first order differential equations, its solution is uniquely determined by the initial condition.
Thus the condition in the r.h.s. of (\ref{onetoone}) is equivalent with that $T^M$ is given by (\ref{TMs}) with $Y^a$ satisfying
\begin{align}
\begin{bmatrix}
Y^1(s_0)\\
Y^2(s_0)
\end{bmatrix}=\sqrt{k\over 2\pi}AV\label{AV}.
\end{align}
Such $Y^a$ is unique for $A$ up to $V$ since the BPS equation is also the set of the first order differential equations.
Therefore, (\ref{TMs}) gives the one to one correspondence (\ref{onetoone}).

If $T^4(s_0)+T^3(s_0)$ is invertible, the condition $T(s_0)=AA^\dg$ can be written explicitly with $T^M(s_0)$ only:\footnote{
Even if $T^4(s_0)+T^3(s_0)$ is not invertible, there is a continuous 
deformation which makes $T^4(s_0)+T^3(s_0)$ invertible.
Concretely, writing $T(s_0)=AA^\dg$, the deformation of this $T(s_0)$ 
into $T(s_0)=A_\epsilon A_\epsilon^\dg$, where
\begin{align}
A_\epsilon=A+\epsilon\cdot
\begin{bmatrix}
1_N\\
0
\end{bmatrix}
\end{align}
with $\epsilon$ continuous parameter, is continuous and makes 
$T^4(s_0)+T^3(s_0)$ invertible if $\epsilon$ is sufficiently small (but nonzero).
Therefore the condition $T(s_0)=AA^\dg$ can be rewritten also for such 
$T(s_0)$, such that there exist a continuous deformation, allowed under 
(\ref{TMconstraint}), to reach that $T(s_0)$.
}
\begin{align}
\begin{cases}
T^4(s_0)+T^3(s_0)\text{ is positive definite},\\
T^4(s_0)-T^3(s_0)=(T^1(s_0)+iT^2(s_0))(T^4(s_0)+T^3(s_0))^{-1}(T^1(s_0)-iT^2(s_0))
\end{cases}.\label{TMconstraint}
\end{align}
The second condition is obtained by writing
\begin{align}
A=
\begin{bmatrix}
A_1\\
BA_1
\end{bmatrix},
\end{align}
where $A_1,B\in Mat_{N\times N}(\mathbf{C})$ and eliminating $B$ from $T(s_0)=AA^\dg$.
Of course, we can have essentially same correspondence between $Y^a$ and $\tilde{T}^M$, instead of $T^M$.

Below we will consider the relation between the $t^M$ and $z^a$, which are $T^M$ and $Y^a$ for $N=1$ 
case and so usual coordinates.
The relation between them is given by
\begin{align}
t^M(z) = {2\pi\over k}\sigma_{ab}^Mz^b\bar{z}^a.
\end{align}
If we parameterize $z^a$ by real coordinates $(r,\theta,\phi,\psi)$ by
\begin{align}
z^1=r\cos\theta e^{i\phi},\nonumber\\
z^2=r\sin\theta e^{i\psi},
\end{align}
where $0\le r<\infty$, \, $0\le\theta\le\pi/2$, \,$\phi\sim\phi+2\pi$, \, $\psi\sim\psi+2\pi$, we have 
\begin{align}
t^1&=R\sin\Theta\cos\Phi,\nonumber\\
t^2&=R\sin\Theta\sin\Phi,\nonumber\\
t^3&=R\cos\Theta,\nonumber\\
t^4&=R,
\end{align}
where
\begin{align}
&0\le R\equiv{4r^2\pi\over k}<\infty,\nonumber\\
&0\le\Theta\equiv2\theta\le\pi,\nonumber\\
&\Phi\equiv\psi-\phi\sim\Phi+2\pi.
\end{align}
Thus, $(t^4)^2=(t^1)^2+(t^2)^2+(t^3)^2$ (which is surely consistent with (\ref{TMconstraint})), and $\{ t^I \}$ 
parameterize ${\mathbf C}^2/U(1)= {\mathbf R}_{\geq 0} \times S^2$, where ${\mathbf Z}_k$ of the 
ABJM action is in the $U(1)$.

The (two set of) extend Nahm data would be related to the D3-NS5 (and D5) system in \cite{ABJM} 
although the space which $T^M$ represent is not flat at least naively.
By now, we should say that a physical meaning or a string theoretical meaning of this map is unclear, however, 
we expect that the correspondence to the (extended) Nahm data will be important for further understanding 
of the M2-M5 brane system.

Here we would like to comment on the relation between our map (\ref{TMs}) and (\ref{tTMs}), and the 
reduction from the M2-branes to the D2-branes discussed by Mukhi and Papageorgakis.
In \cite{Mukhi} they obtained three dimensional Yang-Mills theory from the action of M2-branes by expanding one of the scalars on M2-branes around its vev $v$ and taking the 
$v,k\rightarrow\infty$ limit with $k/v$ fixed.\footnote{
They used the BLG theory, but we can do the same thing also in the ABJM theory.}
For example, if one gives the vev in the $x^3$ direction, the effect of $\z_k$ orbifolding is
\begin{align}
\begin{bmatrix}
x^1+ix^2\\
v+x^3+ix^4
\end{bmatrix}
&\sim
{\rm e}^{i{2\pi\over k}}
\begin{bmatrix}
x^1+ix^2\\
v+x^3+ix^4
\end{bmatrix}\\
&\hspace{-3mm}\underset{k,v\rightarrow\infty}{\rightarrow}
\begin{bmatrix}
x^1+ix^2\\
v+x^3+i(x^4+{2\pi v\over k})
\end{bmatrix}.
\end{align}
Therefore the fluctuation $(x^1,x^2,x^3,x^4)$ lives in $S^1$ times flat $\mathbf{R}^3$.
This $S^1$ becomes so called the M-theory direction and one obtains the D2-D4 bound state in flat spacetime.

Actually, writing 
\begin{align}
Y^1(s)&={k\over 4\pi v}({T^\prime}^1(s)-i{T^\prime}^2(s)),\nonumber\\
Y^2(s)&=v+{k\over 4\pi v}(-{T^\prime}^3(s)+i{T^\prime}^4(s)),\label{MukhiT}
\end{align}
and assuming $k,v\gg |{T^\prime}^I|$, one can obtain the Nahm equation for ${T^\prime}^I$ from the BPS equation (\ref{theBPS}).
For these there is a clear physical interpretation of the system as a D2-D4 bound state
\cite{Mukhi}. 
However, since this procedure 
contains a limit, the information of the BPS solution is considerably lost.
Moreover it can be used only for the BPS solution of the form (\ref{MukhiT}) with $k,v\gg |{T^\prime}^I|$.
On the other hand, our equations for $T^I$ are valid for arbitrary BPS solutions and, together with $T^4$ ($\til{T^4}$), it 
keeps all of the information of $Y^a$ other than $V$.
We also note that, for the BPS solution of the form (\ref{MukhiT}), our $T^I$ and $\til{T}^I$ coincide with ${T^\prime}^I$ 
up to the translation:
\begin{align}
T^{1,2}&=\til{T}^{1,2}+\mathcal{O}\left({1\over k}\right)={T^\prime}^{1,2}+\mathcal{O}\left({1\over k}\right),\\
T^3&=\til{T}^3+\mathcal{O}\left({1\over k}\right)=-{2\pi v^2\over k}+{T^\prime}^3+\mathcal{O}\left({1\over k}\right).\label{coincidence}
\end{align}

\subsection{Examples of the extended Nahm data}

In this subsection, we will show the extended Nahm data $T^M$ and $\tilde{T}^M$ explicitly for the BPS 
solutions obtained in subsection \ref{sfs} and \ref{tH}.
We will compare the parameters of the solutions and the ones of the corresponding Nahm data.
We will see that, in particular, the translation moduli which is trivially realized in the Nahm data is realized non-trivially 
in the solutions in the ABJM.

\subsubsection{The funnel type solutions}
From the funnel type solution (\ref{shifted}), we compute
\begin{align}
T^1(s)&=F_1\tau^1,\nonumber\\
T^2(s)&=F_2\tau^2,\nonumber\\
T^3(s)&=F_3\tau^3+F_4\tau^4,\nonumber\\
T^4(s)&=F_4\tau^3+F_3\tau^4,
\end{align}
where
\begin{align}
F_1(s)&=F_2(s)=f_1f_2=C^2\cdot {\exp[-C^2s/2]\over 1-\exp[-C^2s]},\nonumber\\
F_3(s)&={{f_1}^2+{f_2}^2\over 2}={C^2\over 2}{1+\exp[-C^2s]\over 1-\exp[-C^2s]},\nonumber\\
F_4(s)&={{f_1}^2-{f_2}^2\over 2}=-{C^2\over 2},
\end{align}
and
\begin{align}
\tau^1&={G^1{G^2}^\dg +G^2{G^1}^\dg\over 2},\nonumber\\
\tau^2&=i{G^1{G^2}^\dg -G^2{G^1}^\dg\over 2},\nonumber\\
\tau^3&={G^1{G^1}^\dg -G^2{G^2}^\dg\over 2},\nonumber\\
\tau^4&={G^1{G^1}^\dg +G^2{G^2}^\dg\over 2}.
\end{align}
The matrices $\tau^M$ satisfies
\begin{align}
[\tau^I,\tau^J]&=i\epsilon_{IJK}\tau^K,\\
[\tau^I,\tau^4]&=0,
\end{align}
and then the $\tau^I$ is a generator of the $SU(2)$ as observed in \cite{Nastase:2009ny}.
Explicitly, they are given by
\begin{align}
(\tau^1)_{mn}&={1\over 2}(\delta_{m,n-1}\sqrt{m(N-m-1)}	+\delta_{n,m-1}\sqrt{(m-1)(N-m)}),\nonumber\\
(\tau^2)_{mn}&={i\over 2}(\delta_{m,n-1}\sqrt{m(N-m-1)}-\delta_{n,m-1}\sqrt{(m-1)(N-m)}),\nonumber\\
(\tau^3)_{mn}&=
\begin{cases}
{2m-N\over 2}\delta_{mn}&(m,n<N)\\
0&(m,n=N)
\end{cases},\nonumber\\
(\tau^4)_{mn}&=
\begin{cases}
{N\over 2}\delta_{mn}&(m,n<N)\\
0&(m,n=N)
\end{cases},
\end{align}
where we have used (\ref{Ga}). 
Thus $\tau^I$ is the representation of $({\mathbf N-1}) \oplus {\mathbf 1}$.
In the $s\rightarrow\infty$ limit, the location of the $i$-th D1-brane\footnote{
The D1-brane or the D3-brane are used for the Nahm data $T^I(s)$ 
interpreted as the D1-D3 bound state although we do not know a 
precise relation between this system and the M2-M5 bound state
in the ABJM action considered in this paper.
} 
is
\begin{align}
\begin{bmatrix}
t_i^1\\
t_i^2\\
t_i^3
\end{bmatrix}=
\begin{bmatrix}
0\\
0\\
{C^2(i-N)\over 2}, \,\,
\end{bmatrix}(i=1,2,\cdots N).
\end{align}

One can obtain similar results for $\til{T}^M$.
In that case, coefficient matrices $\til{\tau}^M$, corresponding to $\tau^M$ in $T^M$, are the representation of ${\mathbf N}$ and
\begin{align}
(\til{\tau}^4)_{mn}={N-1\over 2}\delta_{mn}.
\end{align}

\subsubsection{The 't Hooft-Polyakov type solutions}
\begin{figure}[htbp]
 \begin{center}
  \includegraphics[width=100mm]{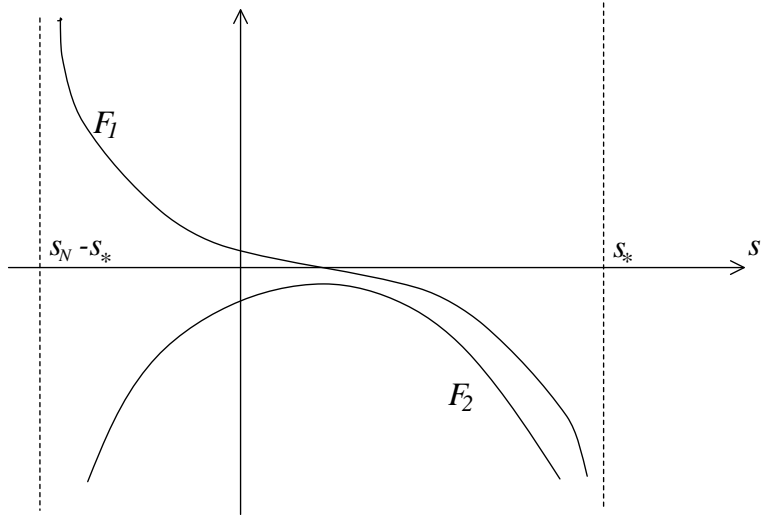}
 \end{center}
\caption{Profiles of $F_1$ and $F_2$ 
\label{fig3}}
\end{figure}

Here we will assume $\gamma\ge\beta\ge\alpha$ for simplicity, since the other cases also give similar results.
The Nahm data obtained from the 't Hooft-Polyakov type solution (\ref{tH1}) is
\begin{align}
T^1(s)&=F_1\cdot{\sigma^1\over 2},\nonumber\\
T^2(s)&=F_2\cdot{\sigma^2\over 2},\nonumber\\
T^3(s)&=F_3\cdot{\sigma^3\over 2}+{\alpha^2-\beta^2-\gamma^2\over 2},\label{2M5TM}
\end{align}
where
\begin{align}
F_1(s)&=2(f_1f_4-f_2f_3),\nonumber\\
F_2(s)&=2(f_3f_1-f_2f_4),\nonumber\\
F_3(s)&=2(f_1f_2-f_3f_4).\label{2M5FI}
\end{align}
This $T^I$ is the Nahm data for the 't Hooft-Polyakov monopole, centered at $(t^1,t^2,t^3)=(0,0,(\alpha^2-\beta^2-\gamma^2)/2)$.
It is well known that the Nahm data for the 't Hooft-Polyakov monopole is explicitly written as
\begin{align}
s_*-s&=\int^{F_1(s)}_{-\infty}{{\mathrm d}F\over\sqrt{(F^2+a)(F^2+b)}},\label{2M5Nahm1}\\
F_2(s)&=-\sqrt{a+(F_1)^2},\nonumber\\
F_3(s)&=-\sqrt{b+(F_1)^2},\label{2M5Nahm23}
\end{align}
where the two parameters
\begin{align}
a&=(F_2)^2-(F_1)^2=4\alpha^2(\gamma^2-\beta^2),\nonumber\\
b&=(F_3)^2-(F_1)^2=4\beta^2(\gamma^2-\alpha^2),
\end{align}
the signs of $F_I$ and the integration constant of (\ref{2M5Nahm1}) are determined by (\ref{2M5FI}) with (\ref{alphabetagamma}), (\ref{theintegrationoff1}) and (\ref{tHooftsolution}).\footnote{
Thus, must should be a mathematical identities relating a bilinear of the elliptic functions and a single elliptic function.
Unfortunately, however, we can not show these identity explicitly here.
}
Using the first kind of elliptic integral (\ref{elliptic}), (\ref{2M5Nahm1}) is written as 
\begin{align}
|s_*-{s_N\over 2}-s|={1\over\sqrt{\max(a,b)}}{\cal F}\left(\arcsin\sqrt{(F_1)^2\over (F_1)^2+\min(a,b)},\sqrt{|a-b|\over\max(a,b)}\right),
\end{align}
where
\begin{align}
s_N={2\over\sqrt{b}}{\cal F}\left(\arcsin(1),\sqrt{(b-a)\over b}\right)
={1\over\sqrt{\beta^2(\gamma^2-\alpha^2)}}{\cal F}\left(\arcsin(1),\sqrt{\gamma^2(\beta^2-\alpha^2)\over\beta^2(\gamma^2-\alpha^2)}\right).\label{sN}
\end{align}
The schematic profiles of $F_1$ and $F_2$ is shown in Fig.~\ref{fig3}.

Now we consider physical interpretation of the solution (\ref{2M5TM}).
The Nahm data contains the translation moduli proportional to the identity matrix with the parameter 
\begin{align}
{\alpha^2-\beta^2-\gamma^2\over 2}.\label{translation}
\end{align}
This is interesting because the shift by the identity matrix of the solution of the BPS equation (\ref{theBPS}) 
does not give another solution in general.
Other than this, there are only two parameters
\begin{align}
a&=4\alpha^2(\gamma^2-\beta^2),\nonumber\\
b&=4\beta^2(\gamma^2-\alpha^2),
\end{align}
which represent the distance between the D3-branes $s_N$ (\ref{sN}) and the ``shape'' of the D1-branes as 
we will see in the next subsection.

\begin{figure}[htbp]
 \begin{center}
  \includegraphics[width=100mm]{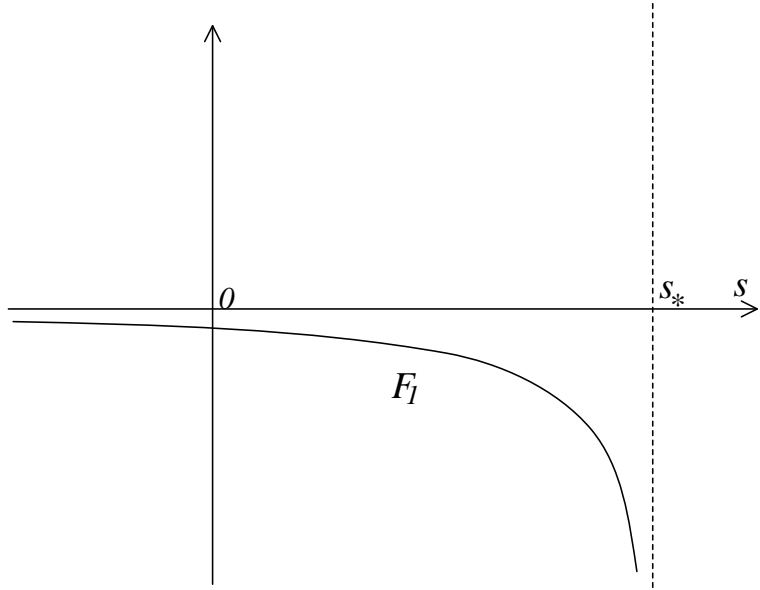}
 \end{center}
\caption{Profile of $F_1$ for $\alpha\le\beta=\gamma$ 
\label{fig5}}
\end{figure}

It is noted that $s_N \ge s_\ast$.
This is seen by comparing (\ref{s*ell}) with $(\alpha_1,\alpha_2,\alpha_3)=(\gamma,\alpha,\beta)$ and (\ref{sN}).
They differ only in the first argument of ${\cal F}$, with which ${\cal F}$ monotonically increases.
Thus, since the first argument of ${\cal F}$ in $s_*$,
\begin{align}
\arcsin\sqrt{1-{\alpha^2\over\gamma^2}},\label{saturation}
\end{align}
is smaller than that in $s_N$, one obtain the inequality.
Indeed, for the solution, from (\ref{2M5FI}), $F_1$ is given by
\begin{align}
F_1(s)&=2\left(f_1\sqrt{f_1^2+\gamma^2}-\sqrt{(f_1^2+\alpha^2)(f_1^2+\beta^2)}\right),
\end{align}
then $F_1(s_*)=-\infty $, while 
\begin{align}
F_1(0)=\gamma^2-\alpha^2-\beta^2,
\end{align}
which is finite because of the cancellation of the $(f_1)^2$ terms.
This means that the locations of the two M5-branes 
are different from the ones of the (hypothetical) D3-branes at least naively.
An extremal case is $\alpha\le\beta=\gamma$, where, despite M2-branes suspending between two finitely 
separated M5 branes, the corresponding D1-branes are attached to a D3-brane and extend to the infinity (Fig.~\ref{fig5}).
This point is interesting and we speculate that, related to the D3-NS5 system, this point would be interpreted naturally.
However, by now, we have not found any concrete interpretation.

On the other hand, $T^4$ is given by
\begin{align}
T^4(s)&=2(f_1f_2+f_3f_4)\cdot{\sigma^3\over 2}+2f_1^2+{\alpha^2+\beta^2+\gamma^2\over 2},
\end{align}
which is divergent at $s=0$ and $s=s_\ast$.
Thus the extended Nahm data should be considered to be defined between the $s=0$ and $s=s_*$.

For the other extended Nahm data $\til{T}^M$, we have 
\begin{align}
\til{T}^1(s)&=2(f_1f_4+f_2f_3)\cdot{\sigma^1\over 2},\nonumber\\
\til{T}^2(s)&=-2(f_3f_1+f_2f_4)\cdot{\sigma^2\over 2},\nonumber\\
\til{T}^3(s)&=-2(f_1f_2+f_3f_4)\cdot{\sigma^3\over 2}+{\alpha^2-\beta^2-\gamma^2\over 2},\nonumber\\
\til{T}^4(s)&=2(f_1f_2-f_3f_4)\cdot{\sigma^3\over 2}-2f_1^2
-{\alpha^2+\beta^2+\gamma^2\over 2},
\end{align}
where the parameters $a$ and $b$ which determine the Nahm data $\til{T}^I=\til{F}^I\sigma^I/2$ are same as 
the ones for $T^I$.
However, $\til{T}^I$ is divergent at $s=0$ and is finite at $s=s_\ast$ in contrast to the profile of $T^I$.

As we said in this section, with the parameters $\alpha,\beta,\gamma$ as (\ref{Mukhiabg}), with $\gamma_0=4\pi v/k$, and taking the $k\rightarrow\infty$ limit, one 
can interpret the Nahm data as actually representing the D2-D4 bound state.
In this case $a,b$ are $a=\alpha_0^2/4\pi^2$ and $b=\beta_0^2/4\pi^2$.
The parameter of the translation (\ref{translation}) is 
$-2\pi v^2/k$, 
so, 
which means, by (\ref{coincidence}), the center of the two D2-branes is at the origin.
Since $\alpha/\gamma=0$, by (\ref{saturation}), the inequality $s_N\ge s_*$ is saturated, and the positions 
of the D4-branes coincide with those of the M5-branes.

\subsection{Conserved quantities by Lax formula}

It is known that the Nahm equation (\ref{eq:TM1}) can be written as the Lax form and then there are infinitely many conserved quantities.
This implies that the BPS equation (\ref{theBPS}) for the M2-M5 bound state also has infinitely many conserved 
quantities because of the map between the solutions of the two sets of the equations shown in this section.

The Nahm equation for the $T^I$ 
is equivalent to the equation in the Lax form: 
\begin{align}
\dot{A}=[A,B],
\end{align}
where
\begin{align}
A(s;\lambda)&={k\over 2\pi}\left(T^3+{\lambda\over 2}(T^1+iT^2)-{1\over 2\lambda}(T^1-iT^2)\right),\\
B(s;\lambda)&=-T^3-\lambda (T^1+iT^2),
\end{align}
for $^\forall\lambda\in\mathbb{C}$.
This enables us to write down the infinitely many conserved quantities \cite{Manton} (which do not need to 
be independent of each other)
\begin{align}
E_n(\lambda)&=\tr(A^n),\\
\dot{E}_n(\lambda)&=0.
\end{align}
In terms of the original variables $Y^a$, we find a simple factorized form:\footnote{
The conserved quantities can be constructed also from $\til{T}^I$ as
\begin{align}	
\til{A}(s;\lambda)&={k\over 2\pi}\left(\til{T}^3+{\lambda\over 2}(\til{T}^1+i\til{T}^2)-{1\over 2\lambda}(\til{T}^1-i\til{T}^2)\right)
=\left({Y^1}^\dg+\lambda{Y^2}^\dg\right)\left(Y^1-{1\over\lambda}Y^2\right),
\end{align}
however, they are not independent of $E_n(\lambda)$ as seen from
\begin{eqnarray}
 \til{E}_n(\lambda)&=\tr(\til{A}^n)=E_n(-1/\lambda).
\end{eqnarray}
}
\begin{align}
A(s;\lambda)&=Y^1{Y^1}^\dg-{1\over\lambda} Y^1{Y^2}^\dg +\lambda Y^2{Y^1}^\dg -Y^2{Y^2}^\dg\nonumber\\
&=\left(Y^1+\lambda Y^2\right)\left({Y^1}^\dg-{1\over\lambda}{Y^2}^\dg\right).
\end{align}

By the symmetry transformation, $E_n(\lambda)$ transforms in the simple way.
Indeed, it is invariant under the $U(N)\times U(N)$.
Under the $SU(2)$ global transformation, $Y^a\rightarrow \Lambda^{ab}$, where 
\begin{align}	
\Lambda=	
\begin{bmatrix}
a &b\\ -\bar{b} &\bar{a}
\end{bmatrix}
\end{align}
and $|a|^2+|b|^2=1$, we have 
\begin{align}
E_n(\lambda)&\rightarrow v^nE_n(\lambda^\prime),\label{su2t}
\end{align}
where
\begin{eqnarray}
v &=& |a|^2-|b|^2-\lambda \bar{a} \bar{b}+{1\over\lambda} ab,\\
\lambda' &=& \frac{b+\lambda \bar{a}}{a-\lambda\bar{b}}.
\end{eqnarray}

We can give some meanings as the D1-D3 brane system to some of the conserved quantities.
The center of the D1-branes may be defined as 
\begin{eqnarray}
\langle t^I \rangle={1\over N}\tr(T^I),
\end{eqnarray}
which is written by $E_1(\lambda)$ as
\begin{align}
\langle t^1 \rangle &={4\pi\over kN}\Re[E_1]_1,\nonumber\\
\langle t^2 \rangle &={4\pi\over kN}\Im[E_1]_1,\nonumber\\
\langle t^3 \rangle &={2\pi\over kN}[E_1]_0,
\end{align}
where $[E_n]_l$ is given by 
\begin{eqnarray}
E_n(\lambda)= \sum_{l \in {\mathbf Z}} [E_n]_l \lambda^l.
\end{eqnarray}
We can also consider the parameters which may represent how the shape of the D1-branes is squashed in the $t^I$ plane: 
\begin{eqnarray}
\delta_I^2={1\over N}\tr((T^I- \langle t^I \rangle )^2.
\end{eqnarray}
These are written by $E_2(\lambda)$ as
\begin{align}
\delta_1^2-\delta_2^2&={16\pi^2\over k^2N}(\Re[E_2]_2)- \langle t^1 \rangle ^2+ \langle t^2 \rangle ^2,\nonumber\\
\delta_2^2-\delta_3^2&={4\pi^2\over k^2N}(-2\Re[E_2]_2-[E_2]_0)- \langle t^2 \rangle ^2+ \langle t^3 \rangle ^2,\nonumber\\
\delta_3^2-\delta_1^2&={4\pi^2\over k^2N}(-2\Re[E_2]_2+[E_2]_0)- \langle t^3 \rangle ^2+ \langle t^1 \rangle ^2.
\end{align}

For the funnel type solutions, 
we obtain
\begin{align}
[A(\lambda)]_{m n} =-{kC^2\over 4\pi} (N-m)\delta_{m n}
\end{align}
by evaluating it at $s\rightarrow\infty$.
Thus, the center of the D1-branes is 
\begin{eqnarray}
\langle t^1 \rangle = \langle t^2 \rangle =0, \,\,\,
\langle t^3 \rangle =-{C^2\over 4}(N-1),
\end{eqnarray}
and the ``shape'' parameters are given by
\begin{align} 
\delta_1^2-\delta_2^2&=0,\nonumber\\
\delta_2^2-\delta_3^2&=-{C^4(N^2-1)\over 48},\nonumber\\
\delta_3^2-\delta_1^2&={C^4(N^2-1)\over 48},
\end{align}
which show that the D1-branes are squashed to the $t^3$ direction.
For the 't Hooft-Polyakov type solutions, we find
\begin{eqnarray}
Y^1+\lambda Y^2=\sqrt{k\over 4\pi}
\begin{bmatrix}
\lambda(\gamma+\beta) &\alpha\\
-\alpha & \lambda(\gamma-\beta)
\end{bmatrix},
\end{eqnarray}
at $s=s_\ast/2$.
Thus we can compute 
\begin{align}
A(\lambda)&=-{k\over 4\pi}\begin{bmatrix}(\gamma+\beta)^2-\alpha^2 &\left(\lambda-{1\over\lambda}\right)\alpha\beta+\left(\lambda+{1\over\lambda}\right)\alpha\gamma\\
\left(\lambda-{1\over\lambda}\right)\alpha\beta-\left(\lambda+{1\over\lambda}\right)\alpha\gamma &(\gamma-\beta)^2-\alpha^2
\end{bmatrix},
\end{align}
and
\begin{eqnarray}
\langle t^1 \rangle = \langle t^2 \rangle =0, \,\,\,
\langle t^3 \rangle =-{\beta^2+\gamma^2-\alpha^2\over 2},
\end{eqnarray}
\begin{align} 
\delta_1^2-\delta_2^2&=\alpha^2(\beta^2-\gamma^2),\nonumber\\
\delta_2^2-\delta_3^2&=\gamma^2(\alpha^2-\beta^2),\nonumber\\
\delta_3^2-\delta_1^2&=\beta^2(\gamma^2-\alpha^2).
\end{align}
In both cases, one can have the solution centered at an arbitrary point by adjusting the parameters of the solution and $SU(2)$ rotation.
Thus we explicitly see that the moduli corresponding to the translation, which is realized trivially in the D2-D4 case, also exists in the M2-M5 case.


\section*{Acknowledgments}

S. T.  would like to thank 
K. Hosomichi for 
useful discussions.
The work of S. T. is partly supported by the Japan Ministry of Education,
Culture, Sports, Science and Technology (MEXT), and by the Grant-in-Aid
for the Global COE program ``The Next Generation of Physics, Spun from
Universality and Emergence'' from the MEXT.

\vspace{1cm}



\end{document}